# From Dirac Notation to Probability Bracket Notation: Time Evolution and Path Integral under Wick Rotations

Xing M. Wang[1]

## Abstract

In this work, we advance the development of the Probability Bracket Notation (PBN), a formalism inspired by Dirac's notation in quantum mechanics, to provide a unified framework for probability modeling. We demonstrate that under a Special Wick Rotation (SWR), an imaginary-time map, the Schrödinger equation, the transition amplitude, and its associated path integral in Dirac notation transform into the master equation, the transition probability, and its Euclidean path integral in the PBN, from which we can reproduce the master equation, representing induced micro-diffusion processes. By extending this approach through a General Wick Rotation (GWR) and employing an anti-Hermitian wave-number operator, we perform parallel derivations of path integrals in both the Dirac and PBN frameworks. This leads to the formulation of the Euclidean Lagrangian for induced diffusions and the strong-damping harmonic oscillator (described by the Smoluchowski diffusion equation). Our findings highlight the versatility of the PBN in bridging quantum mechanics and stochastic processes, offering a coherent notation system for analyzing time evolution and path integrals across these domains.

## 1. Quantum Mechanics and Path Integrals in Dirac Notation

Inspired by the great success of Dirac's Vector Bracket Notation (*VBN*) for vectors in Hilbert spaces, we have proposed the *Probability Bracket Notation* (*PBN*) for probability modeling in probability spaces [1].

In this section, we will give a brief review of Quantum Mechanics (QM) and path integrals in Dirac notation (§1.4 of [2] and §4.3 of [3]). The expressions given here will be used to compare with and shift to those shown in *PBN* later. Readers who are familiar with these Dirac expressions may skip this section.

In Dirac's *VBN*, the inner product of two Hilbert vectors is denoted by a bracket (we call it a v-bracket), which can be split into a v-bra and a v-ket:

$$\text{v-Bracket}: \ \langle\psi_A|\psi_B\rangle \ \Rightarrow \ \text{v-bra}: \ \langle\psi_A|, \quad \text{v-ket}: \ |\psi_B\rangle \tag{1.1}$$

$$\text{where}: \ \langle\psi_A|\psi_B\rangle = \langle\psi_B|\psi_A\rangle^*, \quad \langle\psi_A| = |\psi_A\rangle^\dagger \tag{1.2}$$

***Time-independent Observables and Bases***: Suppose we have a bounded, time-independent standard system ([2], §2.1.1). We begin with three complete orthogonal bases of the Hilbert space formed by three time-independent Hermitian operators: (Hamiltonian, with discrete eigenvalues), (position, with continuous eigenvalues), and (momentum, with continuous eigenvalues). We construct the powerful unit (identity) operators using their eigenvectors as follows:

[1] Sherman Visual Lab, Sunnyvale, CA 94085, USA; xmwang@shermanlab.com

$$\text{Discrete } \varepsilon\text{-basis}: \quad \hat{H}\,|\,\varepsilon_i\rangle = \varepsilon_i\,|\,\varepsilon_i\rangle, \quad \langle\varepsilon_i\,|\,\varepsilon_j\rangle = \delta_{ij}, \quad \hat{I}_\varepsilon = \sum_i |\varepsilon_i\rangle\langle\varepsilon_i\,| \tag{1.3a}$$

$$\text{Continuous } x\text{-basis}: \quad \hat{x}\,|\,x\rangle = x\,|\,x\rangle, \quad \langle x'\,|\,x\rangle = \delta(x-x'), \quad \hat{I}_x = \int dx\,|\,x\rangle\langle x\,| \tag{1.3b}$$

$$\text{Continuous } p\text{-basis}: \quad \hat{p}\,|\,p\rangle = p\,|\,p\rangle, \quad \langle p'\,|\,p\rangle = \delta(p-p'), \quad \hat{I}_p = \int dp\,|\,p\rangle\langle p\,| \tag{1.3c}$$

The stationary system state v-ket can now be expanded by using these unit operators:

$$\text{Discrete } \varepsilon\text{-basis}: \quad |\,\Psi\rangle = \hat{I}_\varepsilon\,|\,\Psi\rangle = \sum_i |\varepsilon_i\rangle\langle\varepsilon_i\,|\,\Psi\rangle \equiv \sum_i |\varepsilon_i\rangle\, c_i \tag{1.4a}$$

$$\text{Continuous x-basis}: \quad |\,\Psi\rangle = \hat{I}_x\,|\,\Psi\rangle = \int dx\,|\,x\rangle\langle x\,|\Psi\rangle \equiv \int dx|\,x\rangle\,\psi(x) \tag{1.4b}$$

$$\text{Continuous } p\text{-basis}: \quad |\,\Psi\rangle = \hat{I}_p\,|\,\Psi\rangle = \int dp\,|\,p\rangle\langle p\,|\Psi\rangle \equiv \int dp|\,p\rangle\, c(p) \tag{1.4c}$$

Similarly, the stationary system state v-bra can be expanded as follows:

$$\langle\Psi\,| = \langle\Psi\,|\,\hat{I}_\varepsilon = \sum\nolimits_i \langle\Psi\,|\varepsilon_i\rangle\langle\varepsilon_i\,| = \sum\nolimits_i \langle\,\varepsilon_i\,|c_i{}^* = |\,\Psi\rangle^\dagger \tag{1.5a}$$

$$\langle\Psi\,| = \langle\Psi\,|\,\hat{I}_x = \int dx\langle\Psi\,|\,x\rangle\langle x\,| = \int dx\langle x\,|\,\psi^*(x) \tag{1.5b}$$

$$\langle\Psi\,| = \langle\Psi\,|\,\hat{I}_p = \int dp\langle\Psi\,|\,p\rangle\langle p\,| = \int dp\langle p\,|\,c^*(p) \tag{1.5c}$$

The expectation value of the three observers can now be easily evaluated by inserting the related unit operator in Eq. (1.3):

$$\langle H\rangle \equiv \overline{H} \equiv \langle\Psi\,|\,\hat{H}\,|\,\Psi\rangle = \sum\nolimits_i \langle\Psi\,|\,\hat{H}\,|\,\varepsilon_i\rangle\langle\varepsilon_i\,|\Psi\rangle \;= \sum\nolimits_i \varepsilon_i\,|\,c_i\,|^2 \tag{1.6a}$$

$$\langle x\rangle \equiv \overline{x} \equiv \langle\Psi\,|\,\hat{x}\,|\,\Psi\rangle = \int \langle\Psi\,|\,\hat{x}\,|\,x\rangle\, dx\langle x\,|\Psi\rangle \;= \int dx\, x\,|\,\psi(x)\,|^2 \tag{1.6b}$$

$$\langle p\rangle \equiv \overline{p} \equiv \langle\Psi\,|\,\hat{p}\,|\,\Psi\rangle = \int \langle\Psi\,|\,\hat{p}\,|\,p\rangle\, dp\langle p\,|\Psi\rangle \;= \int dp\; p\,|\,c(p)\,|^2 \tag{1.6c}$$

Here $|\,c_i\,|^2$ is interpreted as the probability distribution function (PDF) of observable $\hat{H}$, $|\,\psi(x)\,|^2$ is the PDF of observable $\hat{x}$, and $|\,c(p)\,|^2$ is the PDF of observable $\hat{p}$.

***Time evolution in Schrodinger Picture***: The time evolution of state vectors is described by the Schrodinger equation. For a non-relativistic and spin-less particle in one-dimensional time-independent standard form, it reads:

$$i\hbar\frac{\partial}{\partial t}\,|\,\Psi(t)\rangle = \hat{H}(\hat{x},\hat{p})\,|\,\Psi(t)\rangle, \quad \hat{H}(\hat{x},\hat{p}) = \hat{T} + \hat{V} = \frac{1}{2m}\hat{p}^2 + V(\hat{x}) = \hat{H}^\dagger \tag{1.7a}$$

Eq. (1.7a) is independent of representations. We can expand it using the bases of Eq. (1.3):

$$\text{In } \varepsilon\text{-basis:} \quad i\hbar\frac{\partial}{\partial t}\langle\varepsilon_i\,|\,\Psi(t)\rangle = \langle\varepsilon_i\,|\,\hat{H}\,|\,\Psi(t)\rangle \Rightarrow i\hbar\frac{\partial}{\partial t}c_i(t) = \varepsilon_i c_i(t) \tag{1.7b}$$

In $x$-basis: $i\hbar \frac{\partial}{\partial t}\langle x|\Psi(t)\rangle = \langle x|\hat{H}(\hat{x},\hat{p})|\Psi(t)\rangle \Rightarrow i\hbar\frac{\partial}{\partial t}\psi(x,t) = \hat{H}(x,\frac{\hbar}{i}\frac{\partial}{\partial x})\psi(x,t)$ (1.7c)

In $p$-basis: $i\hbar \frac{\partial}{\partial t}\langle p|\Psi(t)\rangle = \langle p|\hat{H}(\hat{x},\hat{p})|\Psi(t)\rangle \Rightarrow i\hbar\frac{\partial}{\partial t}c(p,t) = \hat{H}(i\hbar\frac{\partial}{\partial p},p)c(p,t)$ (1.7d)

Here we have used the following matrix elements of the three operators:

$$H_{ik} = \langle \varepsilon_i|\hat{H}|\varepsilon_k\rangle = \varepsilon_k\delta_{ik} \quad (1.7e)$$

$$\text{In } x\text{-basis: } \langle x|\hat{p}|x'\rangle = \frac{\hbar}{i}\frac{\partial}{\partial x}\delta(x-x') \quad (1.7f)$$

$$\text{In } p\text{-basis: } \langle p|\hat{x}|p'\rangle = i\hbar\frac{\partial}{\partial p}\delta(p-p') \quad (1.7g)$$

Note that all three operators ( $\hat{H},\hat{x}$ and $\hat{p}$ ) are Hermitian operators, hence observables of the quantum system. But, as is well known in QM, we cannot measure two observables simultaneously if they do not commute. For example, we cannot measure at the same time, because $[\hat{x},\hat{p}] = i\hbar \neq 0$ .

The time-dependent Schrodinger equation has a symbolic normalized solution:

$$|\Psi(t)\rangle = \hat{U}(t)|\Psi(0)\rangle, \quad \hat{U}(t) = \exp\left(\frac{-i}{\hbar}\int_0^t \hat{H}(t)dt\right), \quad \hat{U}^\dagger(t) = \hat{U}^{-1}(t) \quad (1.8)$$

Normalization in $\varepsilon$-basis: $1 = \langle\Psi(t)|\Psi(t)\rangle = \sum_i \langle\Psi(t)|\varepsilon_i\rangle\langle\varepsilon_i|\Psi(t)\rangle = \sum_i |c_i(t)|^2$ (1.9a)

Normalization in $x$-basis: $1 = \langle\Psi(t)|\Psi(t)\rangle = \int\langle\Psi(t)|x\rangle dx\langle x|\Psi(t)\rangle = \int dx|\psi(x,t)|^2$ (1.9b)

Normalization in $p$-basis: $1 = \langle\Psi(t)|\Psi(t)\rangle = \int\langle\Psi(t)|p\rangle dp\langle p|\Psi(t)\rangle = \int dp|c(p,t)|^2$ (1.9c)

This is called the *Schrodinger picture*. All three bases are equivalent; we can transform from one to another by a *unitary transformation*. For example, a v-bracket can be transformed from $x$-basis to $p$-basis by using the following *unitary transformation*:

$$\langle x|p\rangle = \psi_p(x) = \frac{1}{\sqrt{2\pi\hbar}}e^{ip/\hbar}, \quad \langle p|x\rangle = \phi_x(p) = \psi^*{}_p(x) = \frac{1}{\sqrt{2\pi\hbar}}e^{-ip/\hbar} \quad (1.10a)$$

They are also the $p$-eigenvectors in $x$-basis and the $x$-eigenvectors in $p$-basis. Using (1.10a), we can transform the state vector from $x$-basis to $p$-basis like:

$$\psi(x,t) = \langle x|\psi(t)\rangle = \langle x|\left\{\int|p\rangle dp\langle p|\right\}|\psi(t)\rangle = \frac{1}{\sqrt{2\pi\hbar}}\int dp\, e^{ip/\hbar}c(p,t) \quad (1.10b)$$

***Heisenberg picture***: Using (1.8), the time-dependent expectation value of a time-independent observable $\hat{O}_S$ can be expressed as:

$$\langle O(t)\rangle \equiv \langle\Psi(t)|\hat{O}_S|\Psi(t)\rangle = \langle\Psi(0)|\hat{U}(t)^{-1}\hat{O}_S\hat{U}(t)|\Psi(0)\rangle \quad (1.11)$$

If the Hamiltonian is time-independent as in Eq. (1.7a) and if the state is an eigenvector of the Hamiltonian, $\hat{H}(\hat{x},\hat{p})|\Psi_E\rangle = E|\Psi_E\rangle$, we have from Eq. (1.8) and (1.11):

$$\langle O(t)\rangle_E = \langle \Psi_E(0)|\exp\left(\frac{i}{\hbar}Et\right)\hat{O}_S\exp\left(\frac{-i}{\hbar}Et\right)|\Psi_E(0)\rangle = \langle O_S\rangle_E = \text{ const.} \tag{1.12}$$

Now we define a time-dependent operator and rewrite Eq. (1.11) to:

$$\hat{O}_H(t) \equiv \hat{U}(t)^{-1}\hat{O}_S\,\hat{U}(t) \tag{1.13a}$$

$$\langle \hat{O}(t)\rangle = \langle \Psi|\hat{O}_H(t)|\Psi\rangle = \langle \Psi(t)|\hat{O}_S|\Psi(t)\rangle \tag{1.13b}$$

This is called the *Heisenberg picture* ([2], §1.8; [3], §11.12), where the state vector is time-independent, but the previously time-independent observable now becomes time-dependent. Applying Eq. (1.13) to the three operators in Eq. (1.7e-g), we have

$$\langle H(t)\rangle = \langle \Psi|\hat{H}_H(t)|\Psi\rangle = \langle \Psi(t)|\hat{H}|\Psi(t)\rangle = \sum\nolimits_i \varepsilon_i|\langle i|\Psi(t)\rangle|^2 = \sum\nolimits_i \varepsilon_i|c(i,t)|^2 \tag{1.14a}$$

$$\langle x(t)\rangle = \langle \Psi|\hat{x}_H(t)|\Psi\rangle = \langle \Psi(t)|\hat{x}|\Psi(t)\rangle = \int dx\,x|\langle x|\Psi(t)\rangle|^2 = \int dx\,x|\psi(x,t)|^2 \tag{1.14b}$$

$$\langle p(t)\rangle = \langle \Psi|\hat{p}_H(t)|\Psi\rangle = \langle \Psi(t)|\hat{p}|\Psi(t)\rangle = \int dp\,p|\langle p|\Psi(t)\rangle|^2 = \int dp\,p|c(p,t)|^2 \tag{1.14c}$$

From Eq. (1.8), the operator in Eq. (1.13a) has the following time-dependent equation:

$$i\hbar\frac{d}{dt}\hat{O}_H(t) = i\hbar\frac{d}{dt}[\hat{U}(t)^\dagger\hat{O}_S\hat{U}(t)] = [\hat{O}_H,\hat{H}] \tag{1.15}$$

***Path integral***: In 1D-QM, the transition amplitude of a particle moving from point $(x_a,t_a)$ to point $(x_b,t_b)$ is defined as (see Eq. (2.1) of [2])[2]:

$$K(b,a) \equiv K(x_b,t_b\,|\,x_a,t_a) \equiv \langle x_b,t_b\,|\,x_a,t_a\rangle \equiv \langle x_b|\hat{U}(t_b,t_a)|x_a\rangle \tag{1.16}$$

Here $\hat{U}(t,t_0)$ is the time evolution operator, given by Eq. (1.8). Starting from Eq. (1.16), one derives Feynman's path integral (see: Eq. (2.54) and (2.60) of Ref. [2]):

$$\langle x_b,t_b\,|\,x_a,t_a\rangle = \int_{x(t_a)=x_a}^{x(t_b)=x_b} Dx\exp\left[\frac{i}{\hbar}S_{cl}(x)\right] = \int_{x(t_a)=x_a}^{x(t_b)=x_b} Dx\exp\left[\frac{i}{\hbar}\int_{t_a}^{t_b}dt\,L_{cl}(x,\dot{x})\right] \tag{1.17}$$

Here $L_{cl}$ is the classical Lagrangian. For a 1D particle, Feynman's standard form (see (2.48) of [2]) of the Hamiltonian and its Lagrangian with time-dependent potential are:

---

[2] In [2], the transition amplitude is denoted by: $K(b,a) \equiv (x_b, t_b\,|\,x_a, t_a)$.

$$H(x,\dot{x},t) = T + V = \frac{m}{2}\dot{x}^2 + V(x,t), \quad H(x,p,t) = \frac{1}{2m}p^2 + V(x,t) \tag{1.18a}$$

$$L_{cl}(x,\dot{x},t) = T - V = \frac{m}{2}\dot{x}^2 - V(x,t) \tag{1.18b}$$

For a *free particle*, $V(x) = 0$, the Schrodinger equation and resulting transition amplitude from Feynman's path integral are (see Eq. (2.125) of Ref. [2]):

$$i\hbar\frac{\partial}{\partial t}\psi(x) = \langle x|\hat{H}|\Psi(t)\rangle = -\frac{\hbar^2}{2m}\frac{\partial^2}{\partial x^2}\psi(x), \quad \langle x|\Psi(t)\rangle = \psi(x) \tag{1.19a}$$

$$\langle x_b, t_b | x_a, t_a\rangle = \sqrt{\frac{m}{2\pi i\hbar(t_b - t_a)}}\exp\left[\frac{im(x_b - x_a)^2}{2\hbar(t_b - t_a)}\right] \tag{1.19b}$$

For a *harmonic oscillator,* we have the following Schrodinger Equation in the $x$-basis:

$$i\hbar\frac{\partial}{\partial t}\langle x|\Psi(t)\rangle = \langle x|\hat{H}|\Psi(t)\rangle = \left(-\frac{\hbar^2}{2m}\frac{\partial^2}{\partial x^2} + \frac{m\omega^2}{2}x^2\right)\langle x|\Psi(t)\rangle \tag{1.20}$$

Its transition amplitude is obtained from Feynman's path integral (see Eq. (2.168) of [2]):

$$\begin{aligned}\langle x_b, t_b | x_a, t_a\rangle &= \sqrt{\frac{m\omega}{2\pi i\hbar\sin\omega(t_b - t_a)}} \\ &\cdot\exp\left\{\frac{im\omega}{2\hbar\sin\omega(t_b - t_a)}[({x_b}^2 + {x_a}^2)\cos\omega(t_b - t_a) - 2x_a x_b]\right\}\end{aligned} \tag{1.21}$$

In the next section, we will briefly review the PBN proposed in Ref. [1] and apply it to describe the expectation values and *time evolution* of continuous-time Markov processes (MP), including homogeneous Markov chains (HMC) [4-6].

## 2. Probability Bracket Notation (PBN) and Master Equations

In our previous article [1], we introduced the symbols of *PBN*. The basic proposition is: the *conditional probability* of event $A$ given evidence $B$ in the sample space $\Omega$, $P(A|B) \in \Re$, can be treated as a *probability bracket* or *P-bracket*, which, similar to Eq. (1.1) for the v-bracket in Dirac notation, can be split into a *P*-bra and a *P*-ket:

$$P\text{-bracket: } P(A|B) \quad \Rightarrow \quad P\text{-bra: } P(A|, \quad P\text{-ket: } |B)\,; \quad \text{Note: } P(A| \neq |A)^\dagger \tag{2.1a}$$

The absolute and conditional probabilities are expressed in *PBN* as follows:

$$P(A) = P(A\,|\,\Omega), \quad P(A\,|\,B) = \frac{P(A \cap B)}{P(B)} = \frac{P(A \cap B\,|\,\Omega)}{P(B\,|\,\Omega)} \tag{2.1b}$$

Thus, $P(A|B) \neq P(B|A)$ if $P(A) \neq P(B)$. Moreover,

$$P(A|\,\Omega) < 1 \text{ if } A \subset \Omega\text{ ; } P(\Omega\,|\,B) = \frac{P(\Omega \cap B)}{P(B)} = \frac{P(B)}{P(B)} = 1, \text{ if } \varnothing \subset B \subseteq \Omega \tag{2.1c}$$

In the probability space $(\Omega, X, P)$ of a *continuous* random variable *X*, we proposed [1] the following properties, resembling those of the *x*-basis in Eq. (1.3b)[3]:

$$\text{For } \forall x \in \Omega: \quad X\,|\,x) = x\,|\,x), \quad P(\Omega\,|\,x) = 1, \quad P: x \mapsto P(x) \equiv P(x\,|\,\Omega), \tag{2.2a}$$

$$P(x\,|\,x') = \delta(x - x'), \quad \int_{x\in\Omega} |x)\,dx\,P(x| = I_x \tag{2.2b}$$

The normalization of the PDF of *X*, $P(x)$, can now be expressed similarly to Eq. (1.9b):

$$P(\Omega\,|\,\Omega) = P(\Omega\,|\,I_x\,|\,\Omega) = \int P(\Omega\,|\,x)\,dx\,P(x\,|\,\Omega) = \int_{x\in\Omega} dx\,P(x\,|\,\Omega) = \int_{x\in\Omega} dx\,P(x) = 1 \tag{2.3}$$

The expectation value of the observable *X* can now also be expanded similarly to Eq. (1.6b):

$$\langle X \rangle \equiv \overline{X} \equiv E(X) = P(\Omega\,|\,X\,|\,\Omega) = \int_{x\in\Omega} P(\Omega\,|\,X\,|\,x)dx\,P(x\,|\,\Omega) = \int_{x\in\Omega} dx\,x\,P(x) \tag{2.4}$$

***Master Equation** and **Euclidean Hamiltonian***: The master equation for Markov processes with *continuous-time* in terms of the probability vector can be written as [7-10]:

$$\frac{\partial}{\partial t}|\,\psi(t)\rangle = \hat{G}(t)\,|\,\psi(t)\rangle, \quad |\,\psi(t)\rangle = \hat{U}(t)\,|\,\psi(0)\rangle = \exp\left(\int_0^t \hat{G}(t)dt\right)\psi(0)\rangle \tag{2.5}$$

The probability vector is equivalent to a *P*-ket in *PBN,* and its time evolution reads:

$$|\,\Psi(t)\rangle \Leftrightarrow |\,\Omega_t): \quad \frac{\partial}{\partial t}|\,\Omega_t) = \hat{G}\,|\,\Omega_t), \quad |\,\Omega_t) = \hat{U}(t)\,|\,\Omega_0) = \exp\left(\int_0^t \hat{G}(t)dt\right)|\,\Omega_0) \tag{2.6}$$

The master equation in *PBN* is also representation-independent, just like the Schrodinger Eq. (1.8) in Dirac notation. The operator $\hat{G}$ is the *Euclidean Hamiltonian*, corresponding to the Hamiltonian operator in Eq. (1.8).

If our probability space is from a *Fock space*, we have the following discrete basis from the *occupation numbers* [1]:

$$\hat{n}_i\,|\,\vec{n}) = n_i\,|\,\vec{n}), \quad P(\vec{n}\,|\,\vec{n}') = \delta_{\vec{n},\vec{n}'} = \prod_i \delta_{n_i,n'_i}, \quad \sum_{\vec{n}} |\vec{n})P(\vec{n}| = I_{\vec{n}} \tag{2.7}$$

[3] The properties of discrete random variables in PBN resemble Eq. (1.3a); for more details, see Ref. [1].

Using the basis in Eq. (2.7), the state *P*-ket, state *P*-bra, and the expectation value of $\hat{F}(\vec{n})$ become:

$$|\Omega_t) = \sum_{\vec{n}} |\vec{n})P(\vec{n}|\Omega_t) \equiv \sum_{\vec{n}} m(\vec{n},t)|\vec{n}), \quad P(\Omega_t| = \sum_{\vec{n}} P(\vec{n}| \tag{2.8a}$$

$$\therefore \langle \hat{F}(\vec{n}) \rangle = P(\Omega_t | F(\hat{\vec{n}}) | \Omega_t) = \sum_{\vec{n},\vec{n}'} P(\vec{n}' | F(\vec{n}) \; m(\vec{n},t) | \vec{n}) = \sum_{\vec{n}} F(\vec{n}) \; m(\vec{n},t) \tag{2.8b}$$

They resample Eq. (1.4a), (1.5a), and (1.6a) in Dirac VBN for a discrete basis. For one-dimensional *Fock space*, we can act on Eq. (2.5) from the left by a *P*-bra $P(n|$ :

$$\frac{\partial}{\partial t} P(n|\Omega_t) = P(n|\hat{G}|\Omega_t) = \sum_m P(n|\hat{G}|m)P(m|\Omega_t) \equiv \sum_m G_{nm} P(m|\Omega_t) \tag{2.9}$$

$$\text{Or:} \quad \frac{\partial}{\partial t} P_n(t) = \sum_m G_{nm} P_m(t), \quad \text{where } P_n(t) \equiv P(n|\Omega_t) \tag{2.10}$$

Eq. (2.9-10) resembles Eq. (1.7b), the Schrodinger Equation in the ε-representation.

The master equation of (2.5) in continuous *x*-basis of (2.2) looks similar to Eq. (1.7c), the Schrodinger equation in the *x*-basis:

$$\frac{\partial}{\partial t} P(x|\Omega_t) = P(x|\hat{G}|I_x|\Omega_t) = \int P(x|\hat{G}|x')dx' P(x'|\Omega_t) \tag{2.11a}$$

$$\text{Or:} \quad \frac{\partial}{\partial t} P(x,t) = \int G(x,x',t)dx' P(x',t) \tag{2.11b}$$

In many cases, $G(x,x') = \hat{G}(x,\partial_{x'},x',t)\delta(x-x')$ , then we can write:

$$\frac{\partial}{\partial t} P(x,t) = \hat{G}(x,\partial_x,t)P(x,t) \tag{2.11c}$$

If a stochastic process $X(t) \equiv X_t$ is an HMC, we can always set $X(0) = 0$ [4-6]. We also have the formal solutions of time-dependent *x*-basis (see §5.2 of [1]):

$$\begin{aligned} &P(X_t = x| = P(x,t| = P(x|\hat{U}(t), \quad |X_t = x) = |x,t) = \hat{U}^{-1}(t)|x) \\ &\therefore P(X_t = x|\Omega_0) = P(x,t|\Omega_0) = P(x|\hat{U}(t)|\Omega_0) = P(x|\Omega_t) = P(x,t) \end{aligned} \tag{2.12a}$$

Then, the evolution of the absolute probability distribution function (PDF) can be derived from the transition probability:

$$\begin{aligned} P(x,t) &= P(x|\Omega_t) = P(X_t = x|\Omega_0) = \int dx_0 P(X_t = x|x_0)P(x_0|\Omega_0) \\ &= \int dx_0 P(X_t = x|X_0 = x_0)P(x_0|\Omega_0) = P(X_t = x|X_0 = 0) = P(x,t|0,0) \end{aligned} \tag{2.13a}$$

Here we have used the initial distribution $\delta(0)$ corresponding to $X(0) = x_0 = 0$ :

$$P(x_0 \,|\, \Omega_0) = P(x = x_0 = 0, t = 0 \,|\, \Omega_0) = \delta(x_0 = 0) = \delta(0) \tag{2.13b}$$

***Langevin Equation and Smoluchowski diffusion equation***: In our later discussion, we will try to find the relation between Schrodinger equation (1.7a) and the master equation (2.5) for a subset of MP: *the diffusion process*. The equation of motion for a general diffusion is described in the form of the Langevin equation (see [7], §5.1-§5.4):

$$\frac{dy}{dt} = A(y,t) + B(y,t)\xi(t) \tag{2.14}$$

Here, $A(y,t)$ and $B(y,t)$ are often referred to as the *drift* and *diffusion* terms, and $\xi(t)$ are chosen to be the Langevin stochastic process (white noise) with the following statistical properties (related to the diffusion coefficient $D$):

$$\langle \xi(t) \rangle = 0, \tag{2.15a}$$
$$\langle \xi(t_1)\, \xi(t_2) \rangle = 2D\delta(t_1 - t_2), \tag{2.15b}$$

The ***Fokker-Planck equation*** [9] can then be written in $x$-basis as:

$$\begin{aligned}
&\frac{\partial}{\partial t} P(x,t) = \hat{G}(x, \partial_x, t)\, P(x,t) \\
&= -\frac{\partial}{\partial x}\left\{\left[A(x,t) + DB(x,t)\frac{\partial B(x,t)}{\partial x}\right]P(x,t)\right\} + D\frac{\partial^2}{\partial x^2}\left[B^2(x,t)P(x,t)\right]
\end{aligned} \tag{2.16}$$

For a Brownian motion in an external potential $V(x,t)$, the *Newton-Langevin equation* becomes (see Eq. (5.17) of [7]):

$$m\frac{d^2 x}{dt^2} = -m\gamma\frac{dx}{dt} - \frac{\partial V}{\partial x} + m\xi(t) \tag{2.17a}$$

The Fokker-Planck equation now becomes the famous *Klein-Kramers equation* (see Eq. (5.20) of [7]):

$$\frac{\partial}{\partial t} P = -\mathrm{v}\frac{\partial P}{\partial x} + V'\frac{\partial P}{\partial x} + \left[\gamma\frac{\partial}{\partial \mathrm{v}}\mathrm{v} + D\frac{\partial^2}{\partial \mathrm{v}^2}\right]P, \quad \text{where:} \quad \mathrm{v} = \dot{x}, \quad V' = \partial_x V \tag{2.17b}$$

For over-damped particles, Eq. (2.17b) leads to *the* ***Smoluchowski diffusion equation*** or ***ordinary Fokker-Planck*** equation (see Eq. (18.288) of [2] and Eq. (5.22) of [7]):

$$\frac{\partial}{\partial t} P(x,t) = \hat{G}\, P(x,t) = \left[D\frac{\partial^2}{\partial x^2} + \frac{1}{m\gamma}V'(x)\right]P(x,t) \tag{2.18a}$$

$$\text{where:} \quad V' = \partial_x V, \quad D = k_B T / (m\gamma) \tag{2.18b}$$

***Heisenberg Picture***: So far, all master equations are displayed in the ***Schrodinger picture***. Similar to Eq. (1.13) and (1.15), we can also introduce the Heisenberg picture in *PBN* (see §5.2 of [1]) for time-independent observers:

$$|\Omega_t) = \hat{U}(t)\,|\,\Omega_0) = e^{\hat{G}t}, \quad \Rightarrow \quad \hat{O}_H(t) = \hat{U}^{-1}(t)\,\hat{O}\,\hat{U}(t) \tag{2.19}$$

$$\frac{d}{dt}\hat{O}_H(t) = \frac{d}{dt}[\hat{U}(t)^{-1}\hat{O}\hat{U}(t)] = [\hat{O}_H, G] \tag{2.20}$$

## 3. Special Wick Rotation, Time Evolution, and Induced Diffusions

It is well known [8] that the imaginary time Schrodinger equation becomes a Diffusion-like equation. Specifically, the free particle Schrodinger equation becomes an Einstein-Brown-like equation. In this section, we will show that, under ***Special Wick Rotation*** (SWR) caused by imaginary time rotation[4], many expressions of §1 in Dirac *VBN* for Hilbert space are naturally *shifted* to expressions of §2 in *PBN* for the probability space.

We will discuss only the following diffusion-like process, which in the *x*-basis reads:

$$\frac{\partial}{\partial t}P(x\,|\,\Omega_t) = \hat{G}(x,-\partial_x)\,|\,\Omega_t) = \left\{D\,\partial_x{}^2 - W(x,-\partial_x,t)\right\}P(x\,|\,\Omega_t)$$

$$\text{Or: } \frac{\partial}{\partial t}P(x,t) = \hat{G}\,P(x,t), \quad \hat{G} = \hat{T}_D + \hat{V}_e = D\,\partial_x{}^2 - \hat{W} \tag{3.1a}$$

$$\text{with} \qquad \hat{V}_e = -W(x,-\partial_x,t) \tag{3.1b}$$

Here, we introduced *Euclidean potential* $\hat{V}_e$ to make the *Euclidean Hamiltonian* $\hat{G}$ more resemble the standard Hamiltonian $\hat{H}$ in Eq. (1.7a) or (1.18a).

To simplify our discussion, we will use the following operators and constants:

$$\hat{k} \equiv \frac{\hat{p}}{\hbar} = \frac{1}{i}\frac{\partial}{\partial x}, \quad \mu_\hbar \equiv \frac{m}{\hbar}, \quad u(x) \equiv \frac{V(x)}{m}, \quad \hat{\kappa} \equiv -i\hat{k} \tag{3.2a}$$

The Hermitian wave-number operator ( $\hat{k}$ ) is well known in QM, and the imaginary operator $\hat{\kappa}$ (see Eq. 3.18 & 4.11 below) is closely related to the operator ( $\hat{k}$ ):

$$\hat{\kappa} \equiv \frac{\hat{k}}{i}, \quad \text{in } x\text{-basis:} \quad P(x\,|\,\hat{\kappa}\,|\,x') = -\frac{\partial}{\partial x}\delta(x-x') \tag{3.2b}$$

With (3.2a), Eq. (1.7a) now reads:

---

[4] For Wick notation and path integrals, see: https://en.wikipedia.org/wiki/Wick_rotation, https://en.wikipedia.org/wiki/Path_integral_formulation.

$$i\frac{\partial}{\partial t}\psi(x,t)=\left\{\hat{T}_\hbar+V_\hbar\right\}\psi(x,t)=\left\{\frac{1}{2\mu_\hbar}\hat{k}^2+\mu_\hbar u(x)\right\}\psi(x,t)\equiv\frac{1}{\hbar}\hat{H}(x,\hat{k})\psi(x,t)$$

$$\text{Or: }\frac{\partial}{i\partial t}\psi(x,t)=-\left\{\hat{T}_\hbar+V_\hbar\right\}\psi(x,t)=\left\{\frac{1}{2\mu_\hbar}\frac{\partial^2}{\partial x^2}-\mu_\hbar u(x)\right\}\psi(x,t), \tag{3.3a}$$

$$\text{where: }\hat{T}_\hbar\equiv\frac{1}{\hbar}\hat{T}=\frac{1}{2\mu_\hbar}\hat{k}^2=\frac{-1}{2\mu_\hbar}\frac{\partial^2}{\partial x^2},\quad V_\hbar\equiv\frac{1}{\hbar}V=\frac{m}{\hbar}\frac{V}{m}=\mu_\hbar u(x) \tag{3.3b}$$

We hide the Planck constant ($\hbar$) in the effective mass ($\mu_\hbar$), so we can easily eliminate it when we shift from the microscopic to the macroscopic world.

The ***induced microscopic diffusion*** is a special case of Eq. (3.1a):

$$\frac{\partial}{\partial t}P(x,t)=\hat{G}(x,\hat{\kappa})P(x,t),\quad \hat{G}=\frac{1}{2\mu_\hbar}\frac{\partial^2}{\partial x^2}-\mu_\hbar u(x)=\frac{\hat{\kappa}^2}{2\mu_\hbar}-\mu_\hbar u(x) \tag{3.4}$$

The ***Special Wick Rotation*** (SWR) caused by imaginary time rotation is defined by [1][11]:

$$\text{SWR: } it\to t,\quad |\psi(t)\rangle\to|\Omega_t),\quad \langle x_b,t_b\,|\,x_a,t_a\rangle\to P(x_b,t_b\,|\,x_a,t_a) \tag{3.5}$$

Under SWR, Schrodinger Eq. (3.3a) is indeed shifted to induced micro diffusion (3.4):

$$\frac{\partial}{i\,\partial t}|\psi(t)\rangle=\frac{-1}{\hbar}\hat{H}(\hat{x},\hat{k})|\psi(t)\rangle\to\frac{\partial}{\partial t}|\Omega_t)=G(\hat{x},\hat{\kappa})|\Omega_t) \tag{3.6a}$$

$$\therefore -\frac{1}{\hbar}\hat{H}(\hat{x},\hat{k})=-\frac{\hat{k}^2}{2\mu_\hbar}-\mu_\hbar u(x)\to\hat{G}(\hat{x},\hat{\kappa})=\frac{1}{2\mu_\hbar}\frac{\partial^2}{\partial x^2}-\mu_\hbar u(x)=\frac{\hat{\kappa}^2}{2\mu_\hbar}-\mu_\hbar u(x) \tag{3.6b}$$

$$\therefore \hat{G}(\hat{x},\hat{\kappa})=-\frac{1}{\hbar}\hat{H}(x,i\hat{\kappa})=\frac{\hat{\kappa}^2}{2\mu_\hbar}-W(x),\quad W(x)=\mu_\hbar u(x)=-V_e(x) \tag{3.6c}$$

Moreover, under SWR (3.5), the path integral of transition amplitude in Eq. (1.16) is now shifted to the path integral of transition probability of an induced micro diffusion as follows:

$$\begin{aligned}&\langle x_b,t_b\,|\,x_a,t_a\rangle=\langle x_b\,|\,\hat{U}(t_b,t_a)\,|\,x_a\rangle=\langle x_b\,|\exp\left\{\int i\,dt\frac{(-1)}{\hbar}\hat{H}(\hat{x},\hat{k})\right\}|\,x_a\rangle\\&\to P(x_b,t_b\,|\,x_a,t_a)=P(x_b\,|\,\hat{U}(t_b,t_a)\,|\,x_a)=P(x_b\,|\exp\left\{\int dt\,\hat{G}(\hat{x},\hat{\kappa})\right\}|\,x_a)\end{aligned} \tag{3.7}$$

This is an expected shift because they share the following properties:

1. Both time evolution operators $\hat{U}(t,t_0)$ for the *Master* equation (2.6) and that for the *Schrodinger* equation (1.8) satisfy the *composition law* (see §1.7 of [2]):

$$\hat{U}(t_n,t_0)=\hat{U}(t_n,t_{n-1})\hat{U}(t_{n-1},t_{n-2})...\hat{U}(t_1,t_0) \tag{3.8}$$

2. For the transition amplitude $K(x_b,t_b\,|\,x_a,t_a)=\langle x_b\,|\,\hat{U}(t_b,t_a)\,|\,x_a\rangle$ in QM, we can insert between each pair of the $\hat{U}$ 's a unit operator from v-bra and v-ket of Eq. (1.3b):

$$\hat{U}(t_i,t_{i-1})\hat{U}(t_{i-1},t_{i-2})=\hat{U}(t_i,t_{i-1})\,\hat{I}_x\,\hat{U}(t_{i-1},t_{i-2}),\quad \hat{I}_x=\int_{-\infty}^{\infty}dx\,|\,x\rangle\langle x\,| \tag{3.9}$$

Similarly, for the transition probability $P(x_b,t_b\,|\,x_a,t_a)=P(x_b\,|\,\hat{U}(t_b,t_a)\,|\,x_a)$, we can also insert between each pair of the $\hat{U}$ 's a unit operator from *P*-bra and *P*-ket of Eq. (2.2b):

$$\hat{U}(t_i,t_{i-1})\hat{U}(t_{i-1},t_{i-2})=\hat{U}(t_i,t_{i-1})\,I_x\,\hat{U}(t_{i-1},t_{i-2}),\quad I_x=\int_{-\infty}^{\infty}dx\,|\,x)\,P(x\,| \tag{3.10}$$

Therefore, under SWR, the Schrodinger equation and its path integral of a conservative system are *simultaneously* shifted to the master equation and its path integral of an *induced micro diffusion*. Now we show two simple but useful examples of path integrals under SWR (3.5-3.7).

***Induced Micro Einstein-Brown motion***: The simplest example is the induced microscopic *Einstein-Brown motion*, when $u(x)=0$ in Eq. (3.3-3.4). Applying (3.5) to Eq. (1.19b) for a free particle, we get the transition probability:

$$P(x_b,t_b\,|\,x_a,t_a)=\sqrt{\frac{1}{4\pi D_\hbar(t_b-t_a)}}\exp\left[-\frac{(x_b-x_a)^2}{4D_\hbar(t_b-t_a)}\right] \tag{3.11}$$

Here we have introduced the *induced micro diffusion coefficient*,

$$D_\hbar\equiv 1/(2\mu_\hbar)=\hbar/(2m) \tag{3.12a}$$

It is related to the *macroscopic* diffusion coefficient $D$, derived from the *Newton-Langevin equation* with potential $V(x)=0$, a viscous force $-m\gamma\dot{x}$ $(m\gamma=6\pi\mu a)$, and a fluctuation force $\xi(t)$ (see Eq. (2.17), (2.18b) or §1.1.2 of [7]):

$$D=\frac{k_BT}{m\gamma}\equiv\frac{1}{2\mu_D} \tag{3.12b}$$

Comparing Eq. (3.12a) and (3.12b), we derive the microscopic friction coefficient:

$$D_\hbar=D\quad\rightarrow\quad\frac{\hbar}{2m}=\frac{k_BT}{m\gamma}\quad\rightarrow\quad\gamma=\frac{2k_BT}{\hbar} \tag{3.13}$$

In the vacuum of outer space, the micro-friction coefficient of induced micro-diffusion can be calculated from Eq. (3.13) $T\approx 2.7°K$ .

Applying Eq. (2.13a) to Eq. (3.11), we have the absolute PDF initially at $x=0$:

$$P(x,t) = P(x\,|\,\Omega_t) = P(x,t\,|\,0,0) = \sqrt{\frac{1}{4\pi D_\hbar\, t}}\exp\left[-\frac{x^2}{4D_\hbar\, t}\right] \tag{3.14}$$

***Induced Micro Diffusion of Harmonic Oscillator***: Our second example is the induced microscopic diffusion of a Harmonic Oscillator. Applying (3.5) to Eq. (1.21), we have the transition probability for harmonic MC:

$$P(x_b,t_b\,|\,x_a,t_a) = \sqrt{\frac{\mu_\hbar \omega}{2\pi i \sin[\omega(t_b - t_a)/i]}}$$
$$\times\exp\left\{\frac{i\mu_\hbar\omega}{2\sin[\omega(t_b-t_a)/i]}[({x_b}^2+{x_a}^2)\cos[\omega(t_b-t_a)/i]-2x_a x_b]\right\} \tag{3.15}$$

Using $\sin(y/i) = (\sinh y)/i$ and $\cos(y/i) = (\cosh y)$, we have:

$$P(x_b,t_b\,|\,x_a,t_a) = \sqrt{\frac{\omega}{4\pi D_\hbar \sinh[\omega(t_b - t_a)]}}$$
$$\times\exp\left\{\frac{-\omega}{4D_\hbar \sinh[\omega(t_b-t_a)]}[({x_b}^2+{x_a}^2)\cosh[\omega(t_b-t_a)]-2x_a x_b]\right\} \tag{3.16}$$

Here $D_\hbar$ is defined in Eq. (3.12a). Following Ref [2], the transition amplitude of the Harmonic oscillator, Eq. (2.168) or our Eq. (1.19), is shifted by an analytic continuation of the time difference $t_b - t_a$ to imaginary values $-i(t_b - t_a)$. The resulting equation in [2], Eq. (2.405), is identical to our Eq. (3.16). It is used to evaluate the quantum statistical partition of harmonic oscillators in contact with a reservoir of temperature $T$ (see Eq. (2.399) of [2]). The partition function is given by the following integral of transition probability (3.16) from "time" $t = 0$ to $t = \hbar\beta \equiv \hbar/(k_B T)$:

$$Z_\omega = \int_{-\infty}^{\infty} dx\, P(x,\hbar\beta\,|\,x,0) = \frac{1}{2\sinh[\beta\hbar\omega/2]}, \quad \beta \equiv \frac{1}{k_B T} \tag{3.17}$$

***The anti-Hermitian wave-number operator*** $\hat{\kappa}$: The operator $\hat{\kappa}$ introduced in Eq. (3.2b) is anti-Hermitian, because:

$$\hat{k} = i\hat{\kappa}, \quad [\hat{x},\hat{k}] = i \to [\hat{x},\hat{\kappa}] = 1, \quad \therefore \hat{k}^\dagger = \hat{k} \to \hat{\kappa}^\dagger = -\hat{\kappa} \tag{3.18}$$

Using this anti-Hermitian operator, the transformation (3.6b) can be expressed as:

$$-\frac{1}{\hbar}\hat{H}(x,\hat{k}) = \frac{-1}{2\mu_\hbar}\hat{k}^2 - \mu_\hbar u(x) \underset{\hat{k}\to i\hat{\kappa}}{\longrightarrow} \hat{G}(x,\hat{\kappa}) = \frac{1}{2\mu_\hbar}\hat{\kappa}^2 - \mu_\hbar u(x) \tag{3.19}$$

The Schrodinger equation of the harmonic oscillator, Eq. (1.20), is now shifted under SWR to an induced micro diffusion as follows:

$$\frac{\partial}{i\,\partial t}\psi(x,t)=\frac{-1}{\hbar}\hat{H}(x,\hat{k})\psi(x,t)=\left\{\frac{-1}{2\mu_{\hbar}}\hat{k}^{2}-\mu_{\hbar}\frac{\omega^{2}}{2}\right\}\psi(x,t) \tag{3.20}$$

$$\to\frac{\partial}{\partial t}P(x,t)=G(x,\hat{\kappa})\,P(x,t)=\left\{\frac{1}{2\mu_{\hbar}}\hat{\kappa}^{2}-\mu_{\hbar}\frac{\omega^{2}}{2}\right\}P(x,t) \tag{3.21}$$

Thus, under Special Wick Rotation, the Schrodinger equation of harmonic oscillator (1.20) or (3.20) and its path integral (1.21) are *simultaneously* shifted to the master equation (3.21) and its path integral solution (3.16) of an induced micro diffusion.

We can also play with operator algebra. In the Heisenberg picture (1.13) of QM, it is well known that (see §11.12 of [3]):

$$\frac{d}{dt}\hat{x}(t)=\frac{d}{dt}\left[U^{-1}(t)\,\hat{x}\,U(t)\right]=\frac{1}{i\hbar}[\hat{x}(t),\hat{H}]=\frac{1}{2mi\hbar}[\hat{x}(t),\hat{p}^{2}]=\frac{\hat{p}(t)}{m}=\frac{\hat{k}(t)}{\mu_{\hbar}} \tag{3.22}$$

We have a similar relation in probability space: using the Heisenberg picture (2.19-20), the evolution equation (2.5), and the commutation relation in Eq. (3.18), we have:

$$\frac{d}{dt}\hat{x}(t)=\frac{d}{dt}\left[U^{-1}(t)\,\hat{x}\,U(t)\right]=[\hat{x}(t),\hat{G}]=\frac{1}{2\mu_{\hbar}}[\hat{x}(t),\hat{\kappa}^{2}]=\frac{\hat{\kappa}(t)}{\mu_{\hbar}} \tag{3.23}$$

It is interesting to investigate induced *Micro Einstein-Brown motion* with constant drift v:

$$P(x,t)=P(x\,|\,\Omega_{t})=\frac{1}{\sqrt{4\pi D_{\hbar}\,t}}\exp\left[-\frac{(x-\mathrm{v}t)^{2}}{4D_{\hbar}\,t}\right],\quad D_{\hbar}=\frac{1}{2\mu_{\hbar}} \tag{3.24}$$

Using Eq. (3.23-24), (3.1b), and (3.12a), we get the real expectation value of "velocity":

$$\langle\dot{x}\rangle=\left\langle\frac{\hat{\kappa}}{\mu_{\hbar}}\right\rangle=P(\Omega\,|\,\frac{\hat{\kappa}}{\mu_{\hbar}}\,|\,\Omega_{t})=\iint dxP(\Omega\,|\,x)\left[-2D_{\hbar}\frac{\partial}{\partial x}\delta(x-x')\right]dx'\,P(x'\,|\,\Omega_{t})$$

$$=\int dx\frac{-(x-\mathrm{v}t)}{t\sqrt{4\pi Dt}}\exp\left[-\frac{(x-\mathrm{v}t)^{2}}{4D_{\hbar}\,t}\right]=\int dx\frac{\mathrm{v}}{\sqrt{4\pi D_{\hbar}\,t}}\exp\left[-\frac{(x-\mathrm{v}t)^{2}}{4D_{\hbar}\,t}\right]=\mathrm{v} \tag{3.25}$$

This suggests that we should use the anti-Hermitian operator $\hat{\kappa}$ instead of the Hermitian operator $\hat{k}$ in Eq. (3.2), which would have given an imaginary expectation value:

$$\langle\dot{x}\rangle=\left\langle\hat{k}\,/\,\mu_{\hbar}\right\rangle=P(\Omega\,|\,i\hat{\kappa}\,/\,\mu_{\hbar}\,|\,\Omega_{t})=i\,\mathrm{v} \tag{3.26}$$

***Implication of Special Wick Rotation***: Our study shows that SWR (3.5) is not just an imaginary time rotation. It also transforms the Schrodinger equation (1.7a) and its path integral (1.16) in Dirac *VBN* to the master equation (3.4) and its path integral (3.7) of an induced microscopic diffusion in *PBN*. According to Eq. (3.8-10), this implies that the *x*-basis (1.3b) in Dirac notation is mapped to the *x*-basis (2.2b) in *PBN* under SWR:

$$\langle x' | x \rangle = \delta(x-x'), \quad \hat{I} = \int dx\, |x\rangle\langle x| \underset{\text{SWR}}{\rightarrow} P(x'|x) = \delta(x-x'), \quad \hat{I} = \int dx\, |x) P(x| \tag{3.27}$$

In Ref. [10], we studied microscopic diffusion processes (or Macroscopic Probability Processes, MPP) of a single free 1D particle and the quantum system of non-interacting identical particles in the Fock space. We reproduced the ideal monatomic gas's internal energy in thermodynamics and the expected number of particles in quantum statistics by connecting time with temperature, the *Wick-Matsubara relation*:

$$it \overset{\text{Wick}}{\rightarrow} t \overset{\text{Matsubara}}{\rightarrow} \hbar / kT \tag{3.28}$$

The Wick-Matsubara relation is consistent with the transformation used in Eq. (3.17)

## 4. General Wick Rotation, Path Integral, and Imaginary Wave Number

In this section, we consider the following shift of time-evolution equations:

$$\begin{aligned} &\frac{\partial}{i\,\partial t}\psi(x,t) = \frac{-1}{\hbar} H(x,k,t)\psi(x,t) = \left\{ \frac{1}{2\mu_\hbar}\frac{\partial^2}{\partial x^2} - V_\hbar(x,t) \right\}\psi(x,t) \\ &\rightarrow \frac{\partial}{\partial t} P(x,t) = \hat{G}\, P(x,t) = (\hat{T}_D + \hat{V}_e)\, P(x,t) = \left\{ \hat{T}_D - W(x,\hat{\kappa},t) \right\} P(x,t) \end{aligned} \tag{4.1a}$$

$$\text{where: } \hat{\kappa} = -\frac{\partial}{\partial x}, \quad \hat{T}_D(\hat{\kappa}) = \frac{1}{2\mu_D}\frac{\partial^2}{\partial x^2} = \frac{1}{2\mu_D}\hat{\kappa}^2, \quad \mu_D \equiv \frac{1}{2D} = \text{const.} \tag{4.1b}$$

We again used Euclidean potential $\hat{V}_e = -W$ as we did in Eq. (3.1b).

This shift is accomplished by following ***General Wick Rotation*** (GWR):

$$\text{SWR:} \quad it \rightarrow t, \quad |\psi(t)\rangle \rightarrow |\Omega_t), \quad \langle x_b, t_b | x_a, t_a \rangle \rightarrow P(x_b, t_b | x_a, t_a) \tag{4.2a}$$

$$\text{Wave Number Shift: } \hat{k} \rightarrow i\hat{\kappa}, k \rightarrow i\kappa, |k\rangle \rightarrow |i\kappa), \langle k| \rightarrow P(i\kappa| \tag{4.2b}$$

$$\text{Diffusion Shift:} \quad \mu_\hbar \rightarrow \mu_D, \quad V_\hbar(x) = \frac{1}{\hbar}V(x) \rightarrow W(x,\hat{\kappa},t) = -V_e(x,\hat{\kappa},t) \tag{4.2c}$$

There is no unified form for Diffusion shift (4.2c). We only discuss the following diffusions:

Induced Micro Diffusion: $$\mu_\hbar = \mu_D, \quad V_\hbar(x) = \mu_\hbar u(x) \to W(x) = \mu_\hbar u(x) \tag{4.3a}$$

Induced Macro Diffusion: $$\mu_\hbar \to \mu_D, \quad V_\hbar(x) = \mu_\hbar u(x) \to W(x) = \mu_D u(x) \tag{4.3b}$$

Strong Damping: $$\mu_\hbar \to \mu_D, \quad V_\hbar(x) \to W(x,\hat{\kappa}) = \frac{-1}{m\gamma}\frac{\partial}{\partial x}V'(x) = \frac{1}{m\gamma}\hat{\kappa}V'(x) \tag{4.3c}$$

***Path Integral for QM and MC, Side by Side*:** To see this more clearly, let us go several steps in the path integral for MC of continuous states, following the path integral process in *QM* (§2.1, [2]). We assume the Hamiltonian takes Feynman's standard form in Eq. (1.18). Its Schrodinger equation in Dirac *VBN* reads:

$$\frac{\partial}{i\,\partial t}|\psi(t)\rangle = \frac{-1}{\hbar}\hat{H}(\hat{x},\hat{k},t)|\psi(t)\rangle = \left\{-T_\hbar(\hat{k}) - V_\hbar(\hat{x},t)\right\}|\psi(t)\rangle \tag{4.4a}$$

Then, under GWR, it is shifted to the following master equation in *PBN*:

$$\to \frac{\partial}{\partial t}|\Omega_t) = \hat{G}(\hat{x},\hat{\kappa},t)|\Omega_t) = \left\{T_D(\hat{\kappa}) - W(x,\hat{\kappa},t)\right\}|\Omega_t) = \left\{\hat{T}_D + \hat{V}_c\right\}|\Omega_t) \tag{4.4b}$$

To begin with, we divide the time interval $[t_a, t_b]$ into $N+1$ small pieces:

$$\Delta t = t_{n+1} - t_n = (t_b - t_a)/(N+1) > 0 \tag{4.5}$$

Applying Eq. (3.8) to (1.16) with (4.5), we have for *QM*:

$$\langle x_b, t_b | x_a, t_a\rangle = \langle x_b|\hat{U}(t_b,t_a)|x_a\rangle = \langle x_b|\hat{U}(t_b,t_N)...\hat{U}(t_n,t_{n-1})...\hat{U}(t_1,t_a)|x_a\rangle \tag{4.6a}$$

Applying Eq. (3.8) to (4.1) with (4.5), we have for *MC*:

$$P(x_b, t_b | x_a, t_a) = P(x_b|\hat{U}(t_b,t_a)|x_a) = P(x_b|\hat{U}(t_b,t_N)...\hat{U}(t_n,t_{n-1})...\hat{U}(t_1,t_a)|x_a) \tag{4.6b}$$

Now we apply the solution of Eq. (1.8) to (4.6a) for *QM*:

$$\langle x_b, t_b | x_a, t_a\rangle = \prod_{n=1}^{N}\int_{-\infty}^{\infty}dx_n\langle x_n|\hat{U}(t_n,t_{n-1})|x_{n-1}\rangle = \prod_{n=1}^{N}\int_{-\infty}^{\infty}dx_n\langle x_n|e^{-\frac{i}{\hbar}\hat{H}\Delta t}|x_{n-1}\rangle \tag{4.7a}$$

Similarly, we apply the solution of Eq. (2.5) to (4.6b) for *MC*:

$$P(x_b, t_b | x_a, t_a) = \prod_{n=1}^{N}\int_{-\infty}^{\infty}dx_n P(x_n|\hat{U}(t_n,t_{n-1})|x_{n-1}) = \prod_{n=1}^{N}\int_{-\infty}^{\infty}dx_n P(x_n|e^{\hat{G}\Delta t}|x_{n-1}) \tag{4.7b}$$

Next, we insert the unit operator (3.9) into each v-bracket in (4.7a) for *QM*:

$$\begin{aligned}
&\langle x_n | e^{-\frac{i}{\hbar}\hat{H}(\hat{p},\hat{x},t_n)\Delta t} | x_{n-1}\rangle = \int_{-\infty}^{\infty} dx \langle x_n | e^{-\frac{i}{\hbar}V(\hat{x},t_n)\Delta t} | x\rangle\langle x | e^{-\frac{i}{\hbar}T(\hat{p})\Delta t} | x_{n-1}\rangle \\
&= \int_{-\infty}^{\infty} dx\, e^{-\frac{i}{\hbar}V(x,t_n)\Delta t} \delta(x - x_n)\langle x | e^{-\frac{i}{\hbar}T(\hat{p})\Delta t} | x_{n-1}\rangle = e^{-\frac{i}{\hbar}V(x,t_n)\Delta t} \langle x_n | e^{-iT_\hbar(\hat{k})\Delta t} | x_{n-1}\rangle
\end{aligned} \tag{4.8a}$$

Similarly, we insert the unit operator (3.10) into each *P*-bracket in (4.7b) for *MC*:

$$\begin{aligned}
&P(x_n | e^{\hat{G}(\hat{\kappa},\hat{x},t_n)\Delta t} | x_{n-1}) = \int_{-\infty}^{\infty} dx\, P(x_n | e^{-W(\hat{x})\Delta t} | x)P(x | e^{T_D(\hat{\kappa})\Delta t} | x_{n-1}) \\
&= \int_{-\infty}^{\infty} dx\, e^{-W(x,t_n)\Delta t} \delta(x - x_n)P(x | e^{T_D(\hat{\kappa})\Delta t} | x_{n-1}) = e^{W(x,t_n)\Delta t} P(x_n | e^{T_D(\hat{\kappa})\Delta t} | x_{n-1})
\end{aligned} \tag{4.8b}$$

To evaluate the v-bracket in (4.8a), we need the unit operator from the momentum operator and its eigenvectors as given in Eq. (1.3c), (1.7e), and (1.10a):

$$\begin{aligned}
&\hat{p} | p\rangle = p | p\rangle, \quad \langle p | p'\rangle = \delta(p - p'), \quad \int dp | p\rangle\langle p | = I \\
&\langle x | \hat{p} | x'\rangle = \frac{\hbar}{i}\frac{\partial}{\partial x}\delta(x - x'), \quad \langle x | p\rangle = \frac{1}{2\pi\hbar} e^{ipx/\hbar}
\end{aligned} \tag{4.9}$$

From Eq. (4.9), we have the properties of the wave-number operator $\hat{k} \equiv \hat{p} / \hbar$:

$$\hat{k} | k\rangle = k | k\rangle, \quad \langle k | k'\rangle = \delta(k - k'), \quad \int dk | k\rangle\langle k | = I \tag{4.10a}$$

$$\langle x | \hat{k} | x'\rangle = \frac{1}{i}\frac{\partial}{\partial x}\delta(x - x'), \quad \langle x | k\rangle = \frac{1}{\sqrt{2\pi}} e^{ikx}, \quad \langle k | x\rangle = \frac{1}{\sqrt{2\pi}} e^{-ikx} \tag{4.10b}$$

To evaluate the *P*-bracket in (4.8b), we need related properties of the anti-Hermitian operator $\hat{\kappa}$ in probability space, which is given by the wave number shift (4.2b). The properties of the operator $\hat{k}$ in Eq. (4.10) are now shifted to the properties of $\hat{\kappa}$ as follows:

$$\text{(4.10a)} \to i\hat{\kappa} | i\kappa) = i\kappa | i\kappa), P(i\kappa | i\kappa') = \delta(i\kappa - i\kappa'), \int id\kappa | i\kappa)P(i\kappa | = I \tag{4.11a}$$

$$\text{(4.10b)} \to P(x | i\hat{\kappa} | x') = \frac{1}{i}\frac{\partial}{\partial x}\delta(x - x'), P(x | i\kappa) = \frac{1}{\sqrt{2\pi}} e^{-\kappa x}, P(i\kappa | x) = \frac{1}{\sqrt{2\pi}} e^{\kappa x} \tag{4.11b}$$

Renaming $| i\kappa) \to | \kappa)$, $P(i\kappa | \to P(\kappa |$ and using the property of the delta function, we get:

$$\hat{\kappa} | \kappa) = \kappa | \kappa), \quad P(\kappa | \kappa') = -i\,\delta(\kappa - \kappa'), \quad \int id\kappa | \kappa)\, P(\kappa | = I \tag{4.12a}$$

$$P(x | \hat{\kappa} | x') = -\frac{\partial}{\partial x}\delta(x - x'), \quad P(x | \kappa) = \frac{1}{\sqrt{2\pi}} e^{-\kappa x}, \quad P(\kappa | x) = \frac{1}{\sqrt{2\pi}} e^{\kappa x} \tag{4.12b}$$

We can verify the properties related to operator $\hat{\kappa}$ in (4.12) as follows:

$$P(x\,|\,\hat{\kappa}\,|\,\kappa)=\int P(x\,|\,\hat{\kappa}\,|\,x')dx'\,P(x'\,|\,\kappa)=\int -\frac{\partial}{\partial x}\delta(x-x')dx'\frac{1}{\sqrt{2\pi}}e^{-\kappa x'}=\kappa\,P(x\,|\,\kappa) \qquad (4.13)$$

$$P(\kappa\,|\,\kappa')=\int P(\kappa\,|\,x)dx\,P(x\,|\,\kappa')=\frac{1}{2\pi}\int dx\,e^{-x(\kappa-\kappa')}$$

$$\underset{imaginary\,\kappa}{\overset{\kappa=-ik,\kappa'=-ik',}{=}}\frac{1}{2\pi}\int dx\,e^{ix(k-k')}=\delta(k'-k)=\delta(i\kappa'-i\kappa)=-i\delta(\kappa'-\kappa) \qquad (4.14a)$$

$$P(x\,|\,x')=\int P(x\,|\,\kappa)\,id\kappa P(\kappa\,|\,x')=\frac{i}{2\pi}\int d\kappa\,e^{-\kappa(x-x')}$$

$$\underset{imaginary\,\kappa}{\overset{\kappa=-ik}{=}}\frac{1}{2\pi}\int dk\,e^{ik(x-x')}=\delta(x-x') \qquad (4.14b)$$

Now let us go back to Eq. (4.8a), using the operator $\hat{k}$ and $\langle x\,|\,k\rangle$ in Eq. (4.10) to evaluate the v-bracket involving $\hat{T}$ :

$$\langle x_n\,|\,e^{-iT_h(\hat{k})\Delta t}\,|\,x_{n-1}\rangle=\int dk_n\langle x_n\,|\,k_n\rangle\langle k_n\,|\,e^{-iT_h(k_n)\Delta t/\hbar}\,|\,x_{n-1}\rangle$$

$$=\int\frac{dk_n}{2\pi}e^{ik_nx_n}e^{-iT_h(k_n)\Delta t}\langle k_n\,|\,x_{n-1}\rangle=\int\frac{dk_n}{2\pi}e^{ik_n(x_n-x_{n-1})\;-iT_h(k_n)\Delta t} \qquad (4.15a)$$

Similarly, we can evaluate the $P$-bracket involving $\hat{T}_D$ in Eq. (4.8b) by using the expression of the unit operator $\hat{\kappa}$ and $P(x\,|\,\kappa)$ in Eq. (4.12):

$$P(x_n\,|\,e^{T_D(\hat{\kappa})\Delta t}\,|\,x_{n-1})=\int id\kappa_nP(x_n\,|\,e^{T_D(\hat{\kappa})\Delta t}\,|\,\kappa_n)P(\kappa_n\,|\,x_{n-1})$$

$$=\int\frac{id\kappa_n}{2\pi}e^{-\kappa_nx_n}e^{T_D(\kappa_n)\Delta t}P(\kappa_n\,|\,x_{n-1})=\int\frac{id\kappa_n}{2\pi}e^{-\kappa_n(x_n-x_{n-1})\;+T_D(\kappa_n)\Delta t} \qquad (4.15b)$$

Inserting (4.15a) back into (4.8a), we obtain the Feynman path integral formula (see Eq. (2.14) of [2]) for the transition amplitude:

$$\langle x_b,t_b\,|\,x_a,t_a\rangle\approx\prod_{n=1}^{N}\left[\int_{-\infty}^{\infty}dx_n\right]\prod_{n=1}^{N+1}\left[\int_{-\infty}^{\infty}\frac{dp_n}{2\pi\hbar}\right]\exp\left\{\frac{i}{\hbar}[p_n(x_n-x_{n-1})-\Delta t\,H(p_n,x_n,t_n)]\right\}$$

$$=\prod_{n=1}^{N}\left[\int_{-\infty}^{\infty}dx_n\right]\prod_{n=1}^{N+1}\left[\int_{-\infty}^{\infty}\frac{dk_n}{2\pi}\right]\exp\left\{i\,k_n(x_n-x_{n-1})+i\Delta t\frac{(-1)}{\hbar}H(\hbar k_n,x_n,t_n)\right\} \qquad (4.16a)$$

Similarly, inserting (4.15b) back into (4.8b), we obtain the following Euclidean path integral for transition probability:

$$P(x_b,t_b\,|\,x_a,t_a)\approx\prod_{n=1}^{N}\left\{\int_{-\infty}^{\infty}dx_n\right\}\prod_{n=1}^{N+1}\left[\int_{-i\infty}^{i\infty}\frac{id\kappa_n}{2\pi}\right]\exp\left\{-\kappa_n(x_n-x_{n-1})+\Delta t\,G(\kappa_n,x_n,t_n)\right\} \qquad (4.16b)$$

Note that we can get (4.16b) from (4.16a) by using transformation (3.19) and (4.14):

$$i\Delta t \to \Delta t, \quad k_n \equiv \frac{p_n}{\hbar} \to i\kappa_n, \quad -\frac{\hat{H}}{\hbar} \to \hat{G} = \hat{T}_D - \hat{W} = \hat{T}_D + \hat{V}_e \tag{4.16c}$$

Our imaginary wave-number operator $\hat{\kappa}$ is anti-Hermitian; it is not an observable but rather an *auxiliary* variable to complete the path integral (4.8b) and (4.15b), which contains a *P*-bracket involving the operator $\hat{T}_D$ with a differential operator $\hat{\kappa}$. Moreover, it does not commute with the coordinate as shown in Eq. (3.18), so they cannot have joint probabilities. The two P-brackets in Eq. (4.12b) should not be interpreted as conditional probabilities, but as intermediate P-brackets for Euclidean path integrals. It serves a similar auxiliary role as the momentum operator $\hat{p}$ used to complete the path integral (4.8a) and (4.15a), which contains a v-bracket involving a differential operator $\hat{p}$.

**Time Evolution Equation from Path Integral**: We can "recover" the Schrodinger equation and master equation from (4.16a) and (4.16b). First, from Eq. (4.16a), we have the following approximation for the transition amplitude (see §2.1.3 of [2]):

$$\langle x_b, t_b \mid x_a, t_a \rangle \approx \int_{-\infty}^{\infty} dx_N \langle x_b, t_b \mid x_N, t_N \rangle \langle x_N, t_N \mid x_a, t_a \rangle, \quad t_N = t_b - \Delta t \tag{4.17a}$$

where

$$\langle x_b, t_b \mid x_N, t_N \rangle \approx \int_{-\infty}^{\infty} \frac{dp_b}{2\pi\hbar} \exp\left\{ \frac{i}{\hbar} [p_b (x_b - x_N) - \Delta t \, H(p_b, x_b, t_b)] \right\}$$

$$= \exp\left\{ -\frac{i}{\hbar} \Delta t \, H(-i\hbar\partial_{x_b}, x_b, t_b) \right\} \int_{-\infty}^{\infty} \frac{dp_b}{2\pi\hbar} \exp\left\{ \frac{i}{\hbar} [p_b (x_b - x_N)] \right\}$$

$$= \exp\left\{ -\frac{i}{\hbar} \Delta t \, H(-i\hbar\partial_{x_b}, x_b, t_b) \right\} \delta(x_b - x_N) \tag{4.18a}$$

Therefore:

$$\langle x_b, t_b \mid x_a, t_a \rangle \approx \exp\left\{ -\frac{i}{\hbar} \Delta t \, H(-i\hbar\partial_{x_b}, x_b, t_b) \right\} \langle x_b, t_b - \Delta t \mid x_a, t_a \rangle \tag{4.19a}$$

$$\therefore \quad \frac{1}{\Delta t} \left[ \langle x_b, t_b \mid x_a, t_a \rangle - \langle x_b, t_b - \Delta t \mid x_a, t_a \rangle \right]$$

$$\approx \frac{1}{\Delta t} \left\{ \exp\left[ -\frac{i}{\hbar} \Delta t \, H(-i\hbar\partial_{x_b}, x_b, t_b) \right] - 1 \right\} \langle x_b, t_b \mid x_a, t_a \rangle \tag{4.20a}$$

In the limit $\Delta t \to 0$, Eq. (4.20a) leads to the Schrodinger equation (1.7a) for transition amplitude and the momentum operator in the *x*-basis:

$$i\hbar\partial_{t_b} \langle x_b, t_b \mid x_a, t_a \rangle = H\left( -i\hbar\partial_{x_b}, x_b, t_b \right) \langle x_b, t_b \mid x_a, t_a \rangle, \quad \hat{p} = -i\hbar\partial_{x_b}$$

Or:

$$-i\partial_{t_b} \langle x_b, t_b \mid x_a, t_a \rangle = (-1/\hbar) H\left( \hbar\hat{k}, x_b, t_b \right) \langle x_b, t_b \mid x_a, t_a \rangle, \quad \hat{k} = -i\partial_{x_b} \tag{4.21a}$$

On the other hand, from Eq. (4.16b), we have a similar expression for the transition *P*-bracket:

$$P(x_b,t_b\,|\,x_a,t_a) \approx \int_{-\infty}^{\infty} dx_N P(x_b,t_b\,|\,x_N,t_N)P(x_N,t_N\,|\,x_a,t_a), \quad t_N = t_b - \Delta t \tag{4.17b}$$

where
$$P(x_b,t_b\,|\,x_N,t_N) \approx \int_{-i\infty}^{i\infty} \frac{id\kappa_b}{2\pi} \exp\left\{-\kappa_b(x_b - x_N) + \Delta t\, G(\kappa_b, x_b, t_b)\right\}$$

$$\underset{i\kappa_b = k_b}{=} \exp\left\{\Delta t\, G\left(-\partial_{x_b}, x_b, t_b\right)\right\} \int_{-\infty}^{\infty} \frac{dk_b}{2\pi} \exp\left\{[-ik_b(x_b - x_N)]\right\}$$

$$= \exp\left\{\Delta t\, G(-\partial_{x_b}, x_b, t_b)\right\} \delta(x_b - x_N) \tag{4.18b}$$

Therefore:
$$P(x_b,t_b\,|\,x_a,t_a) \approx \exp\left\{\Delta t\, G\left(-\partial_{x_b}, x_b, t_b\right)\right\} P(x_b, t_b - \Delta t\,|\,x_a,t_a) \tag{4.19b}$$

$$\therefore \quad \frac{1}{\Delta t}\left[P(x_b,t_b\,|\,x_a,t_a) - P(x_b,t_b - \Delta t\,|\,x_a,t_a)\right]$$

$$\approx \frac{1}{\Delta t}\left\{\exp\left[\Delta t\, G(-\partial_{x_b}, x_b, t_b)\right] - 1\right\} P(x_b,t_b\,|\,x_a,t_a) \tag{4.20b}$$

In the limit $\Delta t \to 0$, Eq. (3.26b) leads to the master equation (4.1a) for transition probability and the imaginary wave-number operator in the *x*-basis:

$$\partial_{t_b} P(x_b,t_b\,|\,x_a,t_a) = G\left(\hat{\kappa}, x_b, t_b\right) P(x_b,t_b\,|\,x_a,t_a), \quad \hat{\kappa} = -\partial_{x_b} \tag{4.21b}$$

Note again that we can get (4.21b) from (4.21a) by using transformation (4.16c).

***Implication of General Wick Rotation***: The properties of the anti-Hermitian operator $\hat{\kappa}$ are given by the shift (4.2b) of GWR. Its properties play a crucial role in performing the Euclidean path integral. This implies that, under GWR (4.2), not only time (related to the energy), but also the wave-number (related to momentum) becomes imaginary:

$$it \to t, \quad k \to i\kappa \tag{4.22}$$

## 5. Euclidean Path Integral and Euclidean Lagrangian in *PBN*

To continue the path integrals in Eq. (4.16a-b), let us consider the following shift from Feynman's standard form (4.4a-b) to ***induced macro diffusion*** under GWR:

$$\frac{\partial}{i\,\partial t}|\psi_t\rangle = \frac{-1}{\hbar}\hat{H}\,|\psi_t\rangle, \quad \frac{-1}{\hbar}\hat{H} = -\hat{T}_\hbar - V_\hbar(x,t) = \frac{-1}{2\mu_\hbar}\hat{k}^2 - \mu_\hbar\, u(x,t) \tag{5.1a}$$

$$\to \quad \frac{\partial}{\partial t}|\,\Omega_t) = \hat{G}\,|\,\Omega_t), \quad \hat{G} = \frac{1}{2\mu_D}\hat{\kappa}^2 - \mu_D u(x) = \hat{T}_D(\hat{\kappa}) - W(x) = \hat{T}_D + V_e \tag{5.1b}$$

Now, in Eq. (4.16a), we can integrate with momentum as follows (see p.97, §2.1.7, [2]):

$$\int_{-\infty}^{\infty}\frac{dp_n}{2\pi\hbar}\exp\left\{\frac{i}{\hbar}[p_n(x_n-x_{n-1})-\Delta t\,\frac{{p_n}^2}{2m}-\Delta t\,V(x_n,t_n)]\right\}$$

$$=\int_{-\infty}^{\infty}\frac{dp_n}{2\pi\hbar}\exp\left\{-\frac{i}{\hbar}\frac{\Delta t}{2m}\left[p_n-m\frac{(x_n-x_{n-1})}{\Delta t}\right]^2+\frac{m}{2}\Delta t\left(\frac{x_n-x_{n-1}}{\Delta t}\right)^2-\Delta t\,V(x_n,t_n)\right\}$$

$$=\sqrt{\frac{m}{2\pi\hbar i\Delta t}}\exp\left\{\frac{m}{2}\Delta t\left(\frac{x_n-x_{n-1}}{\Delta t}\right)^2-\Delta t\,V(x_n,t_n)\right\}\tag{5.2a}$$

Then Eq. (4.16a) becomes:

$$\langle x_b,t_b\mid x_a,t_a\rangle\approx\frac{1}{\sqrt{2\pi\hbar i\Delta t/m}}\prod_{n=1}^{N}\int_{-\infty}^{\infty}\frac{dx_n}{\sqrt{2\pi\hbar i\Delta t/m}}\exp\left\{\frac{i}{\hbar}A_N\right\},$$

$$\text{where:}\quad A_N=\Delta t\sum_{n=1}^{N+1}\left[\frac{m}{2}\left(\frac{x_n-x_{n-1}}{\Delta t}\right)^2-V(x_n,t_n)\right]$$

$$\underset{\Delta t\to 0}{\rightarrow}A[x]=\int_{t_a}^{t_b}dt\,L_c(x,\dot{x}),\quad L_c(x,\dot{x})=\frac{m}{2}\dot{x}^2-V(x,t)\tag{5.3a}$$

Symbolically, it can be written as (see p.99, §2.1.7, [2]):

$$\langle x_b,t_b\mid x_a,t_a\rangle=\int_{x(t_a)=x_a}^{x(t_b)=x_b}Dx\exp\left[\frac{i}{\hbar}S_{cl}(x)\right]=\int_{x(t_a)=x_a}^{x(t_b)=x_b}Dx\exp\left[\frac{i}{\hbar}\int_{t_a}^{t_b}dt\,L_c(x,\dot{x})\right]$$

$$=\int\ Dx\exp\left\{\int_{t_a}^{t_b}idt\ L_\hbar(x,\dot{x})\right\},\quad L_\hbar(x,\dot{x})=\frac{\mu_\hbar}{2}\dot{x}^2-V_\hbar(x)\tag{5.4a}$$

Similarly, in Eq. (4.16b), we can integrate with imaginary wave number as follows:

$$\int_{-\infty i}^{\infty i}i\frac{d\kappa_n}{2\pi}\exp\left\{[-\kappa_n(x_n-x_{n-1})+\Delta t\,\frac{{\kappa_n}^2}{2\mu_D}-\Delta t\,W(x_n,t_n)]\right\}$$

$$\underset{i\kappa_n=k_n}{=}\int_{-\infty}^{\infty}\frac{dk_n}{2\pi}\exp\left\{-\frac{\Delta t}{2\mu_D}\left[k_n-i\mu_D\frac{(x_n-x_{n-1})}{i\Delta t}\right]^2-\frac{\mu_D}{2}\Delta t\left(\frac{x_n-x_{n-1}}{\Delta t}\right)^2-\Delta t\,W(x_n,t_n)\right\}$$

$$=\sqrt{\frac{\mu_D}{2\pi\Delta t}}\exp\left\{-\frac{\mu_D}{2}\Delta t\left(\frac{x_n-x_{n-1}}{\Delta t}\right)^2-\Delta t\,W(x_n,t_n)\right\}\tag{5.2b}$$

***Euclidean Lagrangian***: Now, Eq. (4.16b) becomes a Euclidean path integral:

$$P(x_b,t_b\mid x_a,t_a)\approx\frac{\sqrt{\mu_D}}{\sqrt{2\pi\Delta t}}\prod_{n=1}^{N}\int_{-\infty}^{\infty}\frac{\sqrt{\mu_D}dx_n}{\sqrt{2\pi\Delta t}}\exp\{-A_N\},$$

$$A_N = \Delta t \sum_{n=1}^{N+1} \left[ \frac{\mu_D}{2} \left( \frac{x_n - x_{n-1}}{\Delta t} \right)^2 + W(x_n, t_n) \right]$$

$$\underset{\Delta t \to 0}{\to} A[x] = \int_{t_a}^{t_b} dt L_e(x, \dot{x}), \quad L_e(x, \dot{x}) = \frac{\mu_D}{2} \dot{x}^2 + W(x,t) = \frac{\mu_D}{2} \dot{x}^2 - V_e(x,t) \tag{5.3b}$$

Symbolically, it can be written as the following Euclidean path integral:

$$P(x_b, t_b \mid x_a, t_a) = \int_{x(t_a)=x_a}^{x(t_b)=x_b} Dx \exp\left\{ -\int_{t_a}^{t_b} dt \left[ \frac{\mu_D}{2} \dot{x}^2 - V_e(x,t) \right] \right\} \tag{5.4b}$$

Compare Eq. (4.27b, 5.4b) with (4.27a, 5.4a), we see that we shift from amplitude to probability by making the following maps induced by GWR:

$$it \to t, \quad \dot{x} \to i\dot{x}, \quad \mu_\hbar \to \mu_D, \quad V_\hbar \to W = -V_e, \quad \langle x_b, t_b \mid x_a, t_a \rangle \to P(x_b, t_b \mid x_a, t_a) \tag{5.5a}$$

Moreover, compare Eq. (4.21a, 5.4a) with (4.21b, 5.4b), we have the following cycle:

$$\dot{x} = \frac{p}{m} \underset{\text{QM}}{\to} \frac{\hat{p}}{m} = \frac{\hat{k}}{\mu_\hbar} \underset{\text{SWR}}{\to} \frac{i\hat{\kappa}}{\mu_\hbar} \underset{\text{Macro}}{\to} \frac{i\hat{\kappa}}{\mu_D} \underset{(3.23)}{\to} i\dot{x} \tag{5.5b}$$

At the same time, the Lagrangian in (5.4a) for Schrodinger Equation (5.1a) is shifted by GWR to the *Euclidean Lagrangian* in (5.4b) for induced *macro* diffusion (5.1b):

$$L_\hbar = \frac{\mu_\hbar}{2} \dot{x}^2 - V_\hbar(x) \quad \underset{\text{GWR}}{\to} \quad L_e = \frac{\mu_D}{2} \dot{x}^2 + W(x) = \frac{\mu_D}{2} \dot{x}^2 - V_e(x) \tag{5.5c}$$

Here we see why Euclidean potential $\hat{V}_e = -W$ was introduced in Eq. (3.1b) and (5.1b). Besides, the Euclidean Lagrangian for the induced *micro* diffusion in Eq. (3.4) can be obtained either by using Eq. (4.3a) to set $\mu_\hbar = \mu_D, V_e(x) = -W = -V_\hbar$ in Eq. (5.5c), or by applying rotations $it \to t, \dot{x} \to i\dot{x}$ directly to the transition amplitude Eq. (5.4a):

$$\exp\left\{ \int_{t_a}^{t_b} idt \left[ \frac{\mu_\hbar}{2} \dot{x}^2 - V_\hbar(x) \right] \right\} \to \exp\left\{ -\int_{t_a}^{t_b} dt \left[ \frac{\mu_\hbar}{2} \dot{x}^2 + V_\hbar(x) \right] \right\} \tag{5.5d}$$

This implies that our path integration procedure in *PBN* is self-consistent.

***Macro Einstein-Brown motion***: The simplest example of GWR is the *Einstein-Brown motion* from induced macro diffusion (4.3b) with $V = W = 0$. Applying (5.5a) to Eq. (1.19b) for a free particle, we get:

$$P(x_b,t_b\,|\,x_a,t_a)=\sqrt{\frac{1}{4\pi D(t_b-t_a)}}\exp\left[-\frac{(x_b-x_a)^2}{4D(t_b-t_a)}\right] \tag{5.6}$$

It is identical to Eq. (3.11), but the *diffusion coefficient D* here should be identified from the *Newton-Langevin equation*, with potential $V(x)=0$ in Eq. (2.17a) and (2.18b):

$$m\frac{d^2x}{dt^2}=-m\gamma\frac{dx}{dt}+\xi(t),\quad \langle x\xi\rangle=0 \tag{5.7a}$$

$$\langle m\dot{x}^2/2\rangle=k_BT\equiv\beta \quad\Rightarrow\quad D=k_BT/m\gamma=1/(m\gamma\beta) \tag{5.7b}$$

Applying Eq. (2.13a) to Eq. (5.6), we obtain the absolute PDF initially at $x = 0$:

$$P(x,t)=P(x\,|\,\Omega_t)=P(x,t\,|\,0,0)=\sqrt{\frac{1}{4\pi Dt}}\exp\left[-\frac{x^2}{4Dt}\right] \tag{5.7d}$$

This is identical to Eq. (1.5) in Ref. [7]. Similar to Eq. (3.24-25), we can calculate the expectation value of “velocity” in a macroscopic *Einstein-Brown motion* with constant drift:

$$P(x,t)=P(x\,|\,\Omega_t)=\frac{1}{\sqrt{4\pi Dt}}\exp\left[-\frac{(x-\mathrm{v}t)^2}{4Dt}\right] \tag{5.8a}$$

Using Eq. (3.23), (4.1b), (4.12), and (5.8a), we get a real value as expected:

$$\langle\dot{x}\rangle=P(\Omega\,|\,(\hat{\kappa}/\mu_D)\,|\,\Omega_t)=\iint dxP(\Omega\,|\,x)\left[-2D\,\partial_x\delta(x-x')\right]dx'\,P(x'\,|\,\Omega_t)=\mathrm{v} \tag{5.8b}$$

***Strong Damping Harmonic Oscillator***: Our next example of GWR is the strong damping diffusion in an external potential, as in Eq. (4.3c). The transformation (5.5a) now reads:

$$it\to t,\quad \mu_\hbar=\frac{m}{\hbar}\to\mu_D=\frac{1}{2D},\quad V_\hbar(x)\to W=\hat{\kappa}\left[\frac{1}{m\gamma}V'(x)\right], \tag{5.9}$$

The master equation is the *Smoluchowski diffusion equation* (2.18), not an induced macro diffusion of (5.1b). Hence, we re-integrate Eq. (4.16b) by using Eq. (5.9) as follows:

$$\begin{aligned}&\int_{-\infty i}^{\infty i}\frac{id\kappa_n}{2\pi}\exp\left\{-\kappa_n(x_n-x_{n-1})+\Delta t\frac{\kappa_n^{\;2}}{2\mu_D}-\Delta t\frac{\kappa_n}{m\gamma}V'(x_n)\right\}\\ &\qquad\underset{i\kappa_n=k_n}{=}\int_{-\infty}^{\infty}\frac{dk_n}{2\pi}\exp\left\{ik_n(x_n-x_{n-1})-\Delta t\frac{k_n^{\;2}}{2\mu_D}+\Delta t\frac{ik_n}{m\gamma}V'(x_n)\right\}\end{aligned} \tag{5.10a}$$

$$=\int_{-\infty}^{\infty}\frac{dk_n}{2\pi}\exp\left\{-\frac{\Delta t}{2\mu_D}\left[k_n-i\mu_D\left(\frac{x_n-x_{n-1}}{\Delta t}+\frac{V'(x_n)}{m\gamma}\right)\right]^2\right\}$$

$$\times\exp\left\{-\frac{\mu_D}{2}\Delta t\left(\frac{x_n-x_{n-1}}{\Delta t}+\frac{1}{m\gamma}V'(x_n)\right)^2\right\} \tag{5.10b}$$

$$=\sqrt{\frac{\mu_D}{2\pi\Delta t}}\exp\left\{-\frac{\mu_D\Delta t}{2}\left[\frac{x_n-x_{n-1}}{\Delta t}+\frac{1}{m\gamma}V'(x_n)\right]^2\right\} \tag{5.10c}$$

Compared with Eq. (5.2b) and (5.3b), Eq. (5.10c) gives us the following Euclidean Lagrangian in the limit of $\Delta t\to 0$:

$$L_e(x,\dot{x})=\frac{\mu_D}{2}\left[\dot{x}+\frac{1}{m\gamma}V'(x)\right]^2=\frac{1}{4D}\left[\dot{x}+\frac{1}{m\gamma}V'(x)\right]^2 \tag{5.11}$$

This is almost identical to the Euclidean Lagrangian of strong damping in Ref. [2] (see §18.9.3 of [2]), except that $\dot{x}$ in (5.11) is replaced by a retarded coordinate $\dot{x}^R$. However, the approach in Ref. [2] is different from ours: it begins with the square of the absolute value of the standard Feynman path integral for QM (see §18.8 of [2]):

$$P(x_b,t_b\mid x_a,t_a)=|\langle x_b,t_b\mid x_a,t_a\rangle|^2=\left|\int\mathcal{D}x(t)\exp\left\{\frac{i}{\hbar}\int_{t_a}^{t_b}dt[\frac{m}{2}\dot{x}^2-V(x)]\right\}\right|^2 \tag{5.12}$$

With some assumptions (see §18.8-9 of [2]) and calculations, the Euclidean Lagrangian (5.11) is obtained. Through an auxiliary momentum integration, it leads to the path integral solution of the Smoluchowski diffusion equation (2.18). Conversely, we began with the Smoluchowski equation (2.18) and subsequently applied it to our path integral formulas (4.16b) to derive the Euclidean Lagrangian (5.11).

For a harmonic oscillator, ($V(x)=m\omega^2x^2/2$), the Smoluchowski Eq. (2.18) becomes:

$$\partial_{t_b}P(x_b,t_b\mid x_a,t_a)=\left\{D\frac{\partial^2}{\partial {x_b}^2}+\frac{\omega^2}{\gamma}\left[1+x_b\frac{\partial}{\partial x_b}\right]\right\}P(x_b,t_b\mid x_a,t_a) \tag{5.14a}$$

$$=\left\{D\frac{\partial^2}{\partial {x_b}^2}+\eta\left(\frac{\partial}{\partial x_b}x_b\right)\right\}P(x_b,t_b\mid x_a,t_a),\quad \eta\equiv\frac{\omega^2}{\gamma} \tag{5.14b}$$

The path integral solution of Eq. (5.14) is well-known (see §18.9.3 of [2], Eq. (4.23) of [7], or Eq. (3.120) of [11]):

$$P(x_b,t_b\mid x_a,t_a)=\frac{1}{\sqrt{2\pi\sigma^2}}\exp\left\{-\frac{[x_b-\overline{x}(t_b-t_a)]^2}{2\sigma^2(t_b-t_a)}\right\} \tag{5.15}$$

Here $\bar{x}(t), \sigma^2(t)$ are the averages defined as:

$$\bar{x}(t) = \langle x(t) \rangle = x_a e^{-\eta t}, \ \ \sigma^2(t) = \langle [x(t) - \bar{x}(t)]^2 \rangle = (D/\eta)(1 - e^{-2\eta t}) \tag{5.16}$$

Because this is an HMC, using Eq. (2.13a), we have the following absolute PDF:

$$P(x,t) = P(x,t\,|\,0,0) = \frac{\sqrt{\eta}}{\sqrt{2\pi D(1-e^{-2\eta t})}} \exp\left\{ -\frac{\eta x^2}{2D(1-e^{-2\eta t})} \right\} \tag{5.17}$$

Note: If $V(x) = 0$ in Eq. (2.18), we can get the Euclidean Lagrangian for *Einstein-Brown motion* either from Eq. (5.11) with $V(x) = 0$, or from Eq. (5.5c) with $W(x) = 0$:

$$L_e(x, \dot{x}) = \frac{\mu_D}{2} \dot{x}^2 = \frac{1}{4D} \dot{x}^2 \tag{5.18}$$

Under the limit $\omega = 0 \to \eta = 0$, Eq. (5.17) produces the same PDF of *Einstein-Brown* motion, identical to Eq. (5.7d), as an example of induced macro diffusion:

$$\lim_{\eta \to 0} \frac{\eta}{(1 - e^{-2\eta t})} = \frac{1}{2t}, \quad \therefore \lim_{\eta \to 0} P(x,t) \underset{(5.17)}{=} \frac{1}{\sqrt{4Dt}} \exp\left\{ -\frac{x^2}{4Dt} \right\} \tag{5.19}$$

## Summary

Using the Probability Bracket Notation (PBN) as a unified probability modeling framework, we demonstrated that, under Special Wick Rotation (SWR), the Schrödinger equation and its path integral of a conservative system in Dirac notation were simultaneously shifted to the master equation and its Euclidean path integral of an induced micro diffusion in the PBN. Next, we extended to General Wick Rotation (GWR) and introduced the anti-Hermitian wave number operator. Then, in parallel with the path integral in Dirac notation in Minkowski space, we performed the Euclidean path integral in the PBN step-by-step and derived the Euclidean Lagrangian of induced diffusions and Smoluchowski diffusion. This study reinforces the applicability of PBN as a versatile framework for modeling stochastic processes, with clear connections to quantum mechanics, further highlighting PBN's potential to describe diffusion processes and non-equilibrium dynamics.

In addition to its foundational contributions, PBN has demonstrated practical utility in more complex settings for probability modeling. Applications to static Bayesian networks (transformations between P-bases) and dynamic Bayesian networks (time-evolution of Markov models), which are detailed in our recently updated preprints [12-13], show how PBN can be integrated into modern probabilistic models, offering a powerful tool for reasoning about dependencies and uncertainty in graphical frameworks for machine learning (ML) and artificial intelligence (AI) [14-15]. Furthermore, our published work on non-Hermitian Hamiltonians [1], specifically Eqs. (110-112), provides a broader context for applying PBN to systems with complex spectra, extending its utility to non-equilibrium statistical mechanics and systems with dissipative dynamics.